\documentclass[twocolumn,superscriptaddress,floatfix,preprintnumbers, nofootinbib,hyperref]{revtex4-2} 
\usepackage{graphicx,amsmath,amssymb,amsfonts,bm,color}
\usepackage{here}
\usepackage{physics}
\usepackage{cases}
\usepackage[hidelinks]{hyperref}
\usepackage{natbib}
\begin{document}

\title{Determining the Spin-Analyzing Powers via Invariants of the Spin Correlation Matrices and Probing the Bell Non-Locality at the Lepton Colliders}
\author{Dianwei Wang}
\affiliation{School of Physics, Henan Normal University, Xinxiang 453007, P.R.China}
\author{Xiqing Hao}
\affiliation{School of Physics, Henan Normal University, Xinxiang 453007, P.R.China}
\author{Liwei Liu}
\affiliation{School of Physics, Henan Normal University, Xinxiang 453007, P.R.China}
\author{Lina Wu}
\affiliation{School of Sciences, Xi’an Technological University, Xi’an 710021, P. R. China}
\author{Tianjun Li}
\email[]{Corresponding author: tli@itp.ac.cn}
\affiliation{School of Physics, Henan Normal University, Xinxiang 453007, P.R.China}

\date{\today}

\begin{abstract}
	
We consider the two-fermion $F_a F_b$ productions and decays via one mediator exchange at the $e^+e^-$ collider. With the assumption that the spin is defined via the Lorentz symmetry, or considering the implicit symmetry in the spin density matrix, we prove that the trace  ${\rm Tr} [C]$ of the spin correlation matrix $C$ is an invariant quantity, and is invariant under basis rotations. Thus, for the exchanges of one mediator such as scalar and gauge boson,  we can determine the product of the spin-analyzing powers for $F_a F_b$ via  ${\rm Tr} [C]$, and reconstruct the spin correlation matrix. With the CHSH-Horodecki criterion, we can probe the Bell non-locality, and evade the no-go theorem. To be concrete, we study the Bell non-locality for the $\Lambda \bar \Lambda $ productions and decays at the BESIII experiment. In addition, the invariant ${\rm Tr} [C]$ is a new physics observable to probe the new physics beyond the Standard Model (SM) and study the SM precision measurements.  Moreover, for the scalar exchanges, we discuss the general invariants of the spin correlation matrices and the related phenomenological consequences. 

\end{abstract}
	
\maketitle


{\textbf{Introduction.}---}The foundations for modern physics are quantum mechanics (QM) and special relativity. In QM, quantum entanglement, which is a correlation between the sub-systems of a composite quantum system, is one characteristic feature~\cite{Einstein:1935rr, Schrodinger:2008pyl}. Such correlation might be preserved even if the sub-systems are 
spatially separated. Also, the Bell inequality~\cite{Bell:1964kc, Clauser:1969ny, Horodecki:1995nsk}, which can be satisfied by the Local Hidden Variable Theories (LHVTs), can be violated by QM. This is called Bell non-locality, the other characteristic feature. Over the past decades, both quantum entanglement and Bell non-locality have been confirmed in the low energy systems, for example, photons, atoms, and solid state qubits, etc~\cite{Freedman:1972zza, Clauser:1978ng, Aspect:1981nv, Aspect:1981zz, Aspect:1982fx,Weihs:1998gy,Hagley:1997bob, Genovese:2005nw, Yin:2017ips, BIGBellTest:2018ebd}, and can be applied in the practical fields such as quantum computing~\cite{Jozsa:2002rcj} and quantum communication~\cite{Curty:2003ybi}, etc.

The Standard Model (SM) for particle physics is described by Quantum Field Theory (QFT),
which is based on QM and special relativity. Therefore, we can probe the
quantum entanglement and Bell non-locality in high-energy collider experiments, which are different from the above low energy experiments. Because many entangled states can be produced in the relativistic scattering and decay processes, we can study various systems at the colliders. For example, the top quark pairs,  $W^{\pm}$ and $Z$ gauge boson pairs, $\tau$ lepton pairs, bottom quark pairs, light quark pairs, $\Lambda$ baryons, and the particles with different spins, etc~\cite{Tornqvist:1980af, Privitera:1991nz, Abel:1992kz, Dreiner:1992gt, Benatti:1997fr, Benatti:1999jt, Benatti:1999du, Bertlmann:2001ea,
Go:2003tx, Banerjee:2014vga, Acin:2000cs, Li:2008dk, Baranov:2008zzb, Chen:2013epa, Qian:2020ini, Banerjee:2015mha, Yongram:2013soa, Cervera-Lierta:2017tdt, Afik:2020onf, ATLAS:2023fsd, CMS:2024pts, Fabbrichesi:2021npl, Aoude:2022imd,Afik:2022dgh,Fabbrichesi:2022ovb, Fabbrichesi:2023idl,
Ehataht:2023zzt,Barr:2021zcp,Barr:2022wyq,Aguilar-Saavedra:2022wam,Fabbrichesi:2023cev,Severi:2021cnj,Larkoski:2022lmv,Aguilar-Saavedra:2022uye,Afik:2022kwm,Gong:2021bcp,Aguilar-Saavedra:2023hss,Aguilar-Saavedra:2024fig,White:2024nuc,Han:2024ugl,Fabbrichesi:2025ywl,Ashby-Pickering:2022umy,Aguilar-Saavedra:2022mpg,Altakach:2022ywa,Aoude:2023hxv,Morales:2023gow,Bernal:2023ruk,Bi:2023uop,Dong:2023xiw,Ma:2023yvd,Sakurai:2023nsc,Bernal:2023jba,Han:2023fci,Cheng:2023qmz,Aguilar-Saavedra:2024hwd,Aguilar-Saavedra:2024vpd,Aguilar-Saavedra:2024whi,Duch:2024pwm,Morales:2024jhj,Subba:2024mnl,Maltoni:2024csn,Afik:2025grr,Wu:2024asu,Cheng:2024btk,Gabrielli:2024kbz,Ruzi:2024cbt,Cheng:2024rxi,Wu:2024ovc,Ruzi:2024iqu,Altomonte:2024upf,Fabbrichesi:2024rec,Cheng:2025cuv,Han:2025ewp,Guo:2026yhz,Han:2023fci,Bernal:2024xhm,DelGratta:2025qyp,Goncalves:2025mvl,Ruzi:2025jql,Hong:2025drg,Goncalves:2025xer,Altakach:2022ywa,LoChiatto:2024dmx,Ruzi:2024cbt,Ding:2025mzj,Pei:2025yvr,Pei:2025ito,Cao:2025qua,Cheng:2025zcf,Shi:2016bvo,Shi:2019mlf,Shi:2019kjf, Pei:2025non,
Pei:2026rlh,Pei:2026wfu,Pei:2026khg,Antozzi:2026vdi, Barr:2024djo, Fabbrichesi:2024wcd, Subba:2024aut, Zhang:2025mmm, Lu:2025hwy, vonKuk:2025kbv,
Zhang:2026nwm, Yang:2026uwu, Aguilar-Saavedra:2026rsx, Fang:2026ddi, Arai:2026jtc, Roberts:2026hxr, Nguyen:2026amq}.

However, we cannot conduct the direct spin measurements and select the measurement settings in the current collider experiments. Thus, the spin correlations cannot be directly measured at the colliders. The main idea for the above studies is that the spin correlation can be measured indirectly at the colliders from the appropriate angular correlations between the final states from the intermediate particle decays.  In particular, the bridges between the spin correlations and
angular correlations are the spin-analyzing powers. Therefore, to evade the circular argument or say logical fallacy, we must measure the spin-analyzing powers and make sure that both QM and QFT cannot be assumed in the studies. This is the great challenge for the studies of quantum entanglement and Bell non-locality in the collider experiments. And thus, the no-go theorem was proposed previously~\cite{Abel:1992kz, Dreiner:1992gt} and revisited recently~\cite{Bechtle:2025ugc,Abel:2025skj,Li:2024luk,Low:2025aqq}.

On the other hand, quantum entanglement and Bell non-locality have been studied traditionally in the Quantum Information Theory (QIT). Thus, we can employ the QIT concepts, and propose the new physics observables to probe the new physics beyond the SM and study the SM precision measurements. The important question is what kind of new physics observables can we propose?

In this paper, we consider the two-fermion $F_a F_b$ productions and decays via one mediator exchange at the $e^+e^-$ collider. With the assumption that the spin is defined via the Lorentz symmetry, or considering the implicit symmetry in the spin density matrix, we prove that the trace  ${\rm Tr} [C]$ of the spin correlation matrix $C$ is an invariant quantity, which is independent on the scattering angle. Also, it is invariant under basis rotations of the polarization axes. We show that ${\rm Tr} [C]$ is equal to 1, 1, and $-3$ for the exchanges of gauge boson, CP-even scalar, and CP-odd scalar, respectively. 
Thus, for one mediator exchange, for example, photon, we can determine the product of the spin-analyzing powers for $F_a F_b$ via  ${\rm Tr} [C]$, and reconstruct  
the spin correlation matrix. With the CHSH-Horodecki criterion~\cite{Clauser:1969ny, Horodecki:1995nsk}, we can probe the Bell non-locality, and evade the no-go theorem. To be concrete, we study the Bell non-locality for the $ e^+ e^- \to J/\psi \to \Lambda \bar \Lambda $ process at the BESIII experiment. And we point out that we can study the Bell non-locality for the Higgs to $\tau^+\tau^-$ at the LHC. In addition, the invariant ${\rm Tr} [C]$ is a brand new physics observable to probe the new physics beyond the SM and study the SM precision measurements. Moreover, for the CP-even scalar and CP-odd scalar exchanges, $C_{11}$, $C_{22}$, and $C_{33}$ are all invariants as well, which are independent on the scattering angle. Thus, we can employ the observables $C_{ii}$ and ${\rm Tr} [C]$ to determine the product of the spin-analyzing powers and study the Bell non-locality, and to probe the new physics beyond the SM and study the SM precision measurements. Similarly, we can consider the invariants of the spin density matrix, for instance, trace, determinant, and rank, etc, and study the quantum entanglement and Bell non-locality at the colliders, which will given elsewhere.

{\textbf{Spin Density Matrix and Spin Correlations from Angular Correlations.}---}At the $e^+e^-$ collider, we consider the production of a pair of fermions  $F_a F_b$, {\it i.e.}, $e^{+}e^{-}\rightarrow F_a F_b$, which subsequently decay as follows 
\begin{equation}
F_{a/b} \rightarrow f_{a/b, 1} + f_{a/b, 2} + ... + f_{a/b, N}~.~\,
\end{equation}
\begin{figure}[!ht]
	\centering
	\includegraphics[]{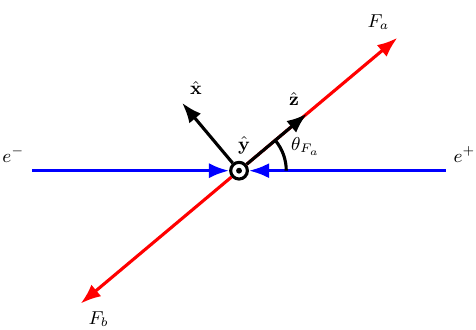}
	\caption{The coordinate system $ (\hat{\vb x}, \hat{\vb y}, \hat{\vb z}) $ for the production process $e^{+}e^{-}\rightarrow F_a F_b$.}
	\label{fig:coordinate}
\end{figure}
We choose the helicity rest frames for $F_a$ and $F_b$, respectively, and define the same coordinate system for them 
\begin{equation}
	\hat{\mathbf{y}}\equiv\frac{\hat{\mathbf{p}}_{e^-}\times\hat{\mathbf{p}}_{F_a}}{\left|\hat{\mathbf{p}}_{e^-}\times\hat{\mathbf{p}}_{F_a}\right|},\ \hat{\mathbf{z}}\equiv \hat{\mathbf{p}}_{F_a},\ \hat{\mathbf{x}}\equiv\hat{\mathbf{y}}\times\hat{\mathbf{z}},\label{eq:helicity-rest-frame}
\end{equation}
as given in FIG.~\ref{fig:coordinate}. This coordinate system is convenient because the difference between the rest frames of $F_a$ and $F_b$ is a pure boost along their momentum directions.
	
We assume that the decay widths of  $F_a$ and $F_b$ are small enough, and thus they can be identified as sharp resonance.
Selecting two fermions $f_{a, 1}$ and $f_{b, 1}$ respectively from the decay products of $F_a$ and $F_b$, we assume that we are inclusive over all final-state momenta except for the unit momentum directions ${\vec q}_{a1}$ and ${\vec q}_{b1}$ for $f_{a, 1}$ and $f_{b, 1}$, respectively. In the following, we shall consider the angular correlations between ${\vec q}_{a1}$ and ${\vec q}_{b1}$, {\it i.e.}, the expectation values $\langle{q^i_{a1}  q^j_{b1}\rangle}$, which can be measured in the collider experiments. And we shall study the relations between 
such angular correlations and the spin correlations for $F_a$ and $F_b$.

The spin polarization state for the composite quantum $F_a F_b$ system is described by a spin density matrix, which can be parametrized as~\cite{Fano:1983zz}
\begin{align}  
	\rho = \frac{I_{4} + \sum_{i}\left(B_{i}^{+}\sigma^{i} \otimes I_{2} + B_{i}^{-}I_{2} \otimes \sigma^{i}\right) + \sum_{i, j} C_{ij} \sigma^{i} \otimes \sigma^{j}}{4} 
	\label{rho1}
\end{align}  
with $i, j = 1, 2, 3$. Here, $I_N$ is the $N\times N$ identity matrix, and
$\sigma_i$ are the Pauli matrices. The coefficients $B_i^{\pm}$ denote  respectively the spin polarizations of $F_a$ and $F_b$, and $C_{ij}$ is the spin correlation matrix. If CP is conserved, we have 
$B_i^{+}=B_i^{-}$ and $ C= C^T$.

Choosing the basis $(|\frac{1}{2}\rangle_{F_a} \otimes |\frac{1}{2}\rangle_{F_b}$, $|\frac{1}{2}\rangle_{F_a} \otimes |-\frac{1}{2}\rangle_{F_b}$, $|-\frac{1}{2}\rangle_{F_a} \otimes |\frac{1}{2}\rangle_{F_b}$, $|-\frac{1}{2}\rangle_{F_a} \otimes |-\frac{1}{2}\rangle_{F_b})$, we can write the spin density matrix as follows~\cite{Werner:1989zz, Horodecki:1997vt}
\begin{align}  
	\rho \equiv \sum^n_{i=1} p_i |\Psi_i \rangle \langle \Psi_i|~,~\sum^n_{i=1} p_i =1 ~,~
	 \label{rho2}
\end{align}
where $n \le 16$, and $p_i$ is a positive real number. Also, $|\Psi_i \rangle \langle \Psi_i|$ is a pure state, and its quantum state (polarization state) $|\Psi_i \rangle$ is defined as~\cite{Pei:2025non}
\begin{align}  
	|\Psi_i \rangle = \sum_{k,j=\pm {1\over 2}}\alpha^i_{k,j} |k\rangle_{F_a} \otimes |j\rangle_{F_b}~,~\,
	\label{psii}
\end{align}  
with the normalization condition  
\begin{align}  
	\sum_{k,j=\pm {1\over 2}}|\alpha^i_{k,j}|^2 = 1~.  
	\label{Normalization}
\end{align}  

From Eq.~(\ref{rho1}), we have
\begin{align}  
	{\rm Tr}[C]=2\left( \rho_{11} +  \rho_{23}+  \rho_{32} + \rho_{44}\right)-1~.  
\end{align}
Thus, from Eqs.~(\ref{rho2}-\ref{Normalization}), we obtain
\begin{align}  
	{\rm Tr}[C]=1-2 \sum^n_{i=1} p_i |\alpha^i_{1/2,-1/2} - \alpha^i_{-1/2,1/2}|^2~.~\,  
	\label{TrC}
\end{align}

Next, we briefly review the relations between the spin correlations $C_{ij}$ and angular correlations $\langle{q^i_{a1}  q^j_{b1}\rangle}$. In the QFT, we can obtain the following relations~\cite{Han:2025ewp, Bechtle:2025ugc} 
\begin{align}  
	C_{ij} = 9\langle{q^i_{a1}  q^j_{b1}\rangle}/\alpha_{F_a}\alpha_{F_b}~,~\,  
	\label{Cqq-1}
\end{align}
where $\alpha_{F_a}$ and $\alpha_{F_a}$ are spin-analyzing powers for $F_a$ and $F_b$, respectively.
In particular, we have $-1\le \alpha_{F_{a/b}} \le 1$.

In the LHVTs, to derive the relations between the spin correlations and angular correlations, we 
cannot assume any concepts from QFT since we test QM against LHVTs.  We briefly review the assumptions  which are needed to establish such relations as follows~\cite{Bechtle:2025ugc, Aguilar-Saavedra:2026rsx}
\begin{enumerate}
	\item A LHVT with a set of local hidden variables.
	\item Special relativity and Poincar\'e-invariance hold.
	\item  The decays of $F_a$ and $F_b$ are independent of each other.
	\item The spin for each particle is an element of reality in the sense of  Einstein-Podolsky-Rosen~\cite{Einstein:1935rr}, {\it i.e.},  a vector with a definite orientation. And thus the spins play the role of hidden variables.
	\item If the particles $F_{a}$ and $F_{b}$ have spins $\hat{s}_{a}$ and $\hat{s}_{b}$, the probability distributions (in their rest frames) of the momenta of the daughter particles $f_{a, 1}$ and $f_{b, 1}$ are always the same and depend only on ${\vec s}_{a}$ and ${\vec s}_{b}$, respectively.
\end{enumerate}
With these assumptions, we can obtain the same relations as given in Eq.~(\ref{Cqq-1}). However, the key difference is that
we now have $-3\le \alpha_{F_{a/b}} \le 3$~\cite{Bechtle:2025ugc}. With such ranges for the spin-analyzing powers, we can show that we can not probe the Bell non-locality at the collider. Therefore, the great challenge is how to measure the product of the spin-analyzing powers $\alpha_{F_a}\alpha_{F_b}$ without the QM and QFT assumptions.

{\textbf{Invariants of the Spin Correlation Matrices.}---}With the assumption that the spin is defined via the Lorentz symmetry or considering the implicit symmetry in the spin density matrix in Eq.~(\ref{rho1}), we will show that we can determine the product of the spin-analyzing powers $\alpha_{F_a}\alpha_{F_b}$ by the invariants of the spin correlation matrices, and probe the Bell non-locality at the lepton collider.

First, we briefly review the definition of spin via the Lorentz symmetry $SO(3,1)$. 
As we know, Poincar\'e symmetry is the basic symmetry for the special relativity, and contains the Lorentz symmetry and translation symmetry. 
There are two Casmir operators for Poincar\'e algebra: $P^{\mu} P_{\mu}$ and $W^{\mu} W_{\mu}$, where $P_{\mu}$ and $W_{\mu}$ are the four-momentum and Pauli–Lubanski pseudo-vector, respectively. These two Casmir operators define the rest mass and spin of a particle, respectively. We consider the proper orthochronous group $SO(3, 1)$ for Lorentz symmetry. The complexified Lorentz algebra for $SO(3, 1)$ is locally isomorphic to the direct sum of two $SU(2)$ algebra, and thus we can formally write it as $SO(3, 1) \simeq SU(2)_1 \times SU(2)_2$. 
Therefore, every finite-dimensional representation of the Lorentz group can be written as two spins $(j_1, j_2)$, where $j_i$ is an integer or half-integer. And then we obtain the representations for the scalar, left-handed Weyl fermion, right-handed Weyl fermion, gauge boson as $(0,0)$, $(1/2,0)$,  $(0,1/2)$,  
$(1/2,1/2)$, respectively. The spin $SU(2)_S$ group can be considered as the diagonal subgroup of  $SU(2)_1 \times SU(2)_2$. 

The above assumption is the solid and fundamental assumption to probe quantum entanglement and Bell non-locality at the colliders. However, it can be relaxed. The key point is the following. The Lorentz symmetry is not violated, and the spin density matrix describes this composite quantum system. And thus, Lorentz symmetry is an implicit symmetry in the spin density matrix. In fact, there exists a $SU(2)_a \times SU(2)_b$ symmetry in the spin density matrix 
in Eq.~(\ref{rho1}). The generators for $SU(2)_a$ and $ SU(2)_b$ are $\sigma^{i} \otimes I_{2}/4$ and $I_{2} \otimes \sigma^{i}/4$, respectively. In particular, $(|\frac{1}{2}\rangle_{F_a}, |-\frac{1}{2}\rangle_{F_a})^T$
forms the fundamental representation of $SU(2)_a$, and $(|\frac{1}{2}\rangle_{F_b}, |-\frac{1}{2}\rangle_{F_b})^T$ forms the fundamental representation of $SU(2)_b$. The spin $SU(2)_S$ group can be considered as the diagonal subgroup of  $SU(2)_a \times SU(2)_b$. 

Because both  $(|\frac{1}{2}\rangle_{F_a}, |-\frac{1}{2}\rangle_{F_a})^T$ and $(|\frac{1}{2}\rangle_{F_b}, |-\frac{1}{2}\rangle_{F_b})^T$ belong to the fundamental representation of $SU(2)_S$, we can decompose the product of these two fundamental representations into the irreducible representations of $SU(2)_S$, {\it i.e.}, decompose the basis of spin density matrix into the irreducible representations of $SU(2)_S$. Such decomposition is
\begin{align}  
\mathbf{2}\otimes \mathbf{2} = \mathbf{1} \oplus \mathbf{3}~,~\,
\end{align}
where the singlet $\mathbf{1}$ and triplet $\mathbf{3}$ belong to the anti-symmetric  and symmetric representations of $SU(2)_S$. In the collider experiments, $\mathbf{1}$ and $\mathbf{3}$ correspond to the CP-odd scalar and gauge boson, respectively, which can couple to both $e^+ e^-$ and $F_a F_b$.

By definition, the quantum state for the singlet in the anti-symmetric representation is 
\begin{align}  
|\Psi_S \rangle = \frac{1}{\sqrt 2} \left(|\frac{1}{2}\rangle_{F_a} \otimes |-\frac{1}{2}\rangle_{F_b} - |-\frac{1}{2}\rangle_{F_a} \otimes |\frac{1}{2}\rangle_{F_b}\right) ~.~\, \nonumber 
\end{align}
And the quantum states for the triplet in the symmetric representation are
\begin{align}  
& |\Psi^{(1,1)}_T \rangle = |\frac{1}{2}\rangle_{F_a} \otimes |\frac{1}{2}\rangle_{F_b} ~,~\, \nonumber \\
& |\Psi^{(1,0)}_T \rangle = \frac{1}{\sqrt 2} \left(|\frac{1}{2}\rangle_{F_a} \otimes |-\frac{1}{2}\rangle_{F_b} + |-\frac{1}{2}\rangle_{F_a} \otimes |\frac{1}{2}\rangle_{F_b}\right) ~,~\, \nonumber \\
& |\Psi^{(1,-1)}_T \rangle = |-\frac{1}{2}\rangle_{F_a} \otimes |-\frac{1}{2}\rangle_{F_b} ~.~\, \nonumber
\end{align}
One can easily prove that these quantum states are orthogonal to each other.

For the spin density matrix in Eq.~(\ref{rho1}), we can obtain the non-zero components for the singlet and triplet
\begin{align} 
& |\Psi_S \rangle:~~ C_{11}=C_{22}=C_{33}=-1; {\rm Tr} [C]=-3~,~\nonumber \\
& |\Psi^{(1,1)}_T \rangle:~~ B^{\pm}_{3}=1, C_{33}=1; {\rm Tr} [C]=1~,~\nonumber \\
& |\Psi^{(1,0)}_T \rangle:~~ C_{11}=C_{22}=-C_{33}=1; {\rm Tr} [C]=1~,~\nonumber \\
& |\Psi^{(1,-1)}_T \rangle:~~B^{\pm}_{3}=-1, C_{33}=1; {\rm Tr} [C]=1~.~\nonumber 
\end{align}	
Thus,  the quantum state for the CP-odd scalar exchange is $|\Psi_S \rangle$, and then we have 
${\rm Tr} [C]=-3$.  Moreover, the quantum states for the gauge boson exchanges are the linear combinations of 
$|\Psi^{(1,1)}_T\rangle$, $|\Psi^{(1,0)}_T\rangle$, and $|\Psi^{(1,-1)}_T\rangle$, whose coefficients are functions of $\theta_{F_a}$. And thus we have ${\rm Tr} [C]=1$.
In addition, the quantum state for the CP-even scalar exchange is  $|\Psi^{(1,0)}_T \rangle$, so we  have
 ${\rm Tr} [C]=1$.

Next, let us prove that  ${\rm Tr} [C]=1$ for the spin correlation matrix via the gauge boson exchange explicitly. The spin density matrix is given in Eq.~(\ref{rho2}), where $|\Psi_i \rangle$ is given 
in Eq.~(\ref{psii}). Because $|\Psi_i \rangle$ and $|\Psi_S \rangle$ are orthogonal to each other,
we obtain 
\begin{align} 
\alpha^i_{1/2,-1/2} = \alpha^i_{-1/2,1/2}~.~
\end{align}	
From Eq.~(\ref{TrC}), we obtain
\begin{align}  
	{\rm Tr}[C]=1-2 \sum^n_{i=1} p_i |\alpha^i_{1/2,-1/2} - \alpha^i_{-1/2,1/2}|^2=1~.~\,  
\end{align}

Interestingly, ${\rm Tr} [C]$ is invariant under the basis rotation of quantization axes. Assuming that
$\vec{e}_i$ and $\vec{e}'_i$ with $i=1,2,3$ are two  different quantization axis choices, which relate to each other via a $SO(3)$ rotation $R_{ij}$, {\it i.e.}, $\vec{e}'_i = \vec{e}_j R_{ji}$, we obtain~\cite{Cheng:2023qmz, Cheng:2024btk}
\begin{align}  
	C'_{ij} = R^T_{ik} C_{kl} R_{lj}=\left(R^TCR\right)_{ij} ~.~\,
\end{align}
Thus, we prove that
\begin{align}  
	{\rm Tr}[C'] = 	{\rm Tr}[R^TCR]= {\rm Tr}[C]~.~\,
\end{align}

In short, for the two-fermion $F_a F_b$ productions and decays via one mediator exchange at the $e^+e^-$ collider, we show that the trace ${\rm Tr} [C]$ of the spin correlation matrix is an invariant quantity,  
which is independent on the scattering angle. In particular,  ${\rm Tr} [C]$ is equal to 1, 1, 
and $-3$ for the exchanges of gauge boson, CP-even scalar, and CP-odd scalar, respectively. 
Thus, for one mediator exchange, for example, photon, we can determine the product of the spin-analyzing powers for $F_a F_b$ via  ${\rm Tr} [C]$, and reconstruct  
the spin correlation matrix. With the CHSH-Horodecki criterion~\cite{Clauser:1969ny, Horodecki:1995nsk}, we can probe the Bell non-locality, and evade the no-go theorem.
In addition, the invariant ${\rm Tr} [C]$ is a new physics observable to probe the new physics beyond the SM and study the SM precision measurements.  
Moreover, for the CP-even scalar and CP-odd scalar changes, $C_{11}$, $C_{22}$, and $C_{33}$ are all invariants as well, which are independent on the scattering angle. In particular,  $C_{33}=-1$ for both CP-even and CP-odd scalars, and then we do not need to know the CP property of the scalar.
Thus, we can employ the observables $C_{ii}$ and ${\rm Tr} [C]$ to determine the product of the spin-analyzing powers and study the Bell non-locality, and to probe the new physics beyond the SM and study the SM precision measurements. 
Similarly, we can study the Bell non-locality for the Higgs to $\tau^+\tau^-$ at the LHC.

By the way, we can perform the calculations in the traditional helicity formalism~\cite{Murayama:1992gi}. And then the decomposition is $\mathbf{2}\otimes \mathbf{\overline{2}} = \mathbf{1} \oplus \mathbf{3}$, where the singlet $\mathbf{1}$ and triplet $\mathbf{3}$ belong to the trivial and adjoint representations of $SU(2)_S$. Although the expressions for the quantum states are different, the physics results are the same.

{\textbf{Bell Non-Locality at the BESIII Experiment.}---}As an example, we study the Bell non-locality for the $\Lambda \bar{\Lambda}$ pair production 
via the $e^+ e^- \rightarrow J/\psi \rightarrow \Lambda \bar{\Lambda}$ process
at the BESIII experiment. The subsequent decays are: $\Lambda\to p+\pi^-$, and $\bar{\Lambda}\to \bar{p}+\pi^+$. So we have
 $F_{a}\equiv \Lambda$ and $F_{b}\equiv \bar{\Lambda}$, and choose $f_{a,1}\equiv p$ and $f_{b,1}\equiv {\bar p}$.
With ${\rm Tr} [C]=1$, we obtain
\begin{equation}
\expval*{q^k_{p} q^k_{\bar p}} = \alpha_{\Lambda} \alpha_{\bar \Lambda}/9~.~
\end{equation}
And thus we have the spin correlation matrix with
\begin{equation}
	C_{ij} = \expval*{q^i_{p} q^j_{\bar p}}/\expval*{q^k_{p} q^k_{\bar p}}~.~
\end{equation}
 Thus, the spin correlation matrix can be determined by the angular correlation measurements.
 
 We denote the eigenvalues of $C^TC$ as $M_1$, $M_2$, $M_3$, which satisfy 
 $M_1 \ge M_2 \ge M_3$. The Bell variable is defined as~\cite{Clauser:1969ny, Horodecki:1995nsk} 
 \begin{equation}
 	\mathcal{B} = 2 \sqrt{M_1+M_2}~.~
 \end{equation}
If $2 < \mathcal{B} \le 2{\sqrt2}$, we realize the Bell non-locality via the CHSH-Horodecki criterion~\cite{Clauser:1969ny, Horodecki:1995nsk}.
\begin{figure}[!ht]
	\centering
	\includegraphics[]{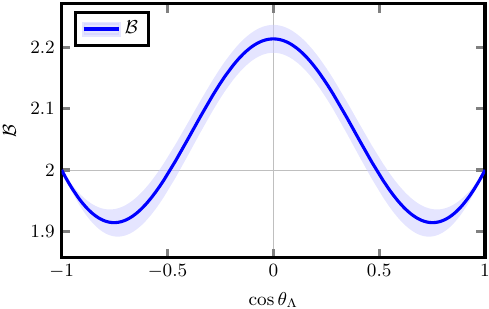}
	\caption{Bell variable vs $\cos\theta_\Lambda$. The shaded region represents the $ 5\sigma $ uncertainty in $ (\alpha_\psi, \Delta\Phi) $ parameter space.}
	\label{fig:criterion}
\end{figure}

Because we do not have the BESIII experimental data, we consider the spin correlation matrix from the QFT calculations~\cite{Perotti:2018wxm, Batozskaya:2023rek, Wu:2024asu} 
\begin{align}  
	\label{eq:Cmn}
&	C_{ij}
	= \frac{1}{1+\alpha_\psi\cos^2\theta_\Lambda}
	\times \nonumber \\
&	\mqty(
	\sin^2\theta_\Lambda&
	0&
	\gamma_\psi\sin\theta_\Lambda\cos\theta_\Lambda\\
	0&
	-\alpha_\psi\sin^2\theta_\Lambda&
	0\\
	\gamma_\psi\sin\theta_\Lambda\cos\theta_\Lambda&
	0&
	\alpha_\psi + \cos^2\theta_\Lambda)~,~\nonumber
\end{align}
where $\alpha_{\psi}\in[-1,+1]$ is the decay parameter of the vector charmonium
$\psi$, and $ \gamma_{\psi} \equiv \sqrt{1-\alpha_{\psi}^{2}}\cos(\Delta\Phi)$
with $\Delta\Phi\in(-\pi,+\pi]$ the relative form factor phase. 

With $\alpha_{\Lambda}= 0.7519 \pm 0.0036 \pm 0.0024$, $\alpha_{\bar \Lambda}= -0.7559 \pm 0.0036 \pm 0.0030 $, $\alpha_\psi =  0.4748 \pm 0.0022 \pm 0.0031$, and $\Delta\Phi = 0.7521 \pm 0.0042 \pm 0.0066 $~\cite{BESIII:2022qax}, we present 
the numerical results in FIG.~\ref{fig:criterion} with solid line. Within $ 5\sigma $ uncertainty in $ (\alpha_\psi, \Delta\Phi) $ parameter space, the Bell non-locality is realized 
in range $ |\cos\theta_\Lambda| < 0.44150 $. 

{\textbf{Conclusion.}---} With the assumption that the spin is defined via the Lorentz symmetry, or considering the implicit symmetry in the spin density matrix, we prove that the trace  ${\rm Tr} [C]$ of the spin correlation matrix $C$ is an invariant quantity, and is invariant under basis rotations. Thus, for the exchanges of one mediator such as scalar and gauge boson, we can determine the product of the spin-analyzing powers via  ${\rm Tr} [C]$, and reconstruct the spin correlation matrix. With the CHSH-Horodecki criterion, we can probe the Bell non-locality, and evade the no-go theorem. As an example, we study the Bell non-locality for the $\Lambda \bar \Lambda $ productions and decays at the BESIII experiment. In particular, the invariant ${\rm Tr} [C]$ is a new physics observable to probe the new physics beyond the SM and study the SM precision measurements.  In addition, for the scalar exchanges, we discuss the general invariants of the spin correlation matrices as well as the related phenomenological consequences.

{\textbf{Acknowledgements.}---}
L Wu is supported in part by the Natural Science Basic Research Program of Shaanxi, Grant No. 2024JC-YBMS-039.
TL is supported in part by the National Key Research and Development Program of China Grant No. 2020YFC2201504, by the Projects No. 11875062, No. 11947302, No. 12047503, and No. 12275333 supported by the National Natural Science Foundation of China, by the Key Research Program of the Chinese Academy of Sciences, Grant No. XDPB15, by the Scientific Instrument Developing Project of the Chinese Academy of Sciences, Grant No. YJKYYQ20190049, by the International Partnership Program of Chinese Academy of Sciences for Grand Challenges, Grant No. 112311KYSB20210012, and by the Henan Province Outstanding Foreign Scientist Studio Project, No.GZS2025008. This work was supported by the High Performance Computing Platform of Henan Normal University.


\bibliography{main-2.bib}

\end{document}